\documentstyle[preprint,aps,eqsecnum,tighten]{revtex}
\begin{document}
\preprint{\baselineskip 18pt{\vbox{\hbox{SU-4240-617}\hbox{UFIFT-HEP-95-15}
\hbox{hep-th/9512047} \hbox{February,1996}}}}
\title{Edge States and Entanglement Entropy}
\vspace{15mm}
\author{A.P. Balachandran$^1$, L. Chandar$^2$ and  Arshad Momen$^1$}
\vspace{15mm}
\address{ $^1$ Department of Physics, Syracuse University,\\
Syracuse, NY 13244-1130, U.S.A.}
\vspace{15mm}
\address{$^2$ Department of Physics, University of Florida,\\
Gainesville, FL 32611, U.S.A.}
\maketitle
\begin{abstract}
It is known that gauge fields defined on manifolds with spatial boundaries
support states localized at the boundaries. In this paper, we demonstrate how
coarse-graining over these states can lead to an entanglement entropy.
In particular, we show that the entanglement entropy of the ground state for
the quantum Hall effect on a disk exhibits  an approximate ``area " law.
\end{abstract}
\vspace{5mm}

\newcommand{\be}{\begin{equation}}
\newcommand{\ee}{\end{equation}}
\newcommand{\bea}{\begin{eqnarray}}
\newcommand{\eea}{\end{eqnarray}}
\newcommand{\real}{{\rm l}\! {\rm R}}
\newcommand{\ra}{\rightarrow}
\newcommand{\tr}{{\rm tr}\;}
\section{Introduction}

Recently there has been renewed interest in the origin of
the so-called area law
for black hole entropy.  One reason for this interest is that it can shed some
light on the possible
substructure for a quantum theory of gravity. In this regard one often invokes
the hypothesis that there are excitations of the black hole horizon leading to
surface states associated to the horizon. In a previous paper \cite{edgr},
using
a very simple treatment of the constraints of canonical gravity, we have shown
that these
states arise in a manner similar to the edge states in quantum Hall effect
(QHE) \cite{23,dhe,jerry}.
We also speculated there that tracing over these edge states might lead
to an entropy for black holes as happens with the calculation of entanglement
entropy \cite{rafael,entropy}.

A complete calculation to verify the above speculation is a rather formidable
task. However, as a step forward, we can try to find evidence in other models
where edge states arise and lead to important physics.
With this in mind, in this paper we study the effective field theory describing
QHE \cite{hall} on a disk and calculate the
entanglement entropy \cite{rafael,entropy} that arises when we trace out the edge
degrees of freedom.
More specifically, we see that the ground state of the
system turns out to have non-trivial correlations between the bulk and the edge
degrees of freedom. For
 this reason, tracing over the edge degrees of
freedom leads to an impure density matrix.  The entanglement
entropy that we calculate is the entropy associated with this density matrix.
The result we find is quite interesting
 - the entaglement entropy obeys an
approximate area law [ that is the entropy is approximately
proportional to the perimeter of the disk ]
for weak coupling between the edge and the bulk.  An area
law of the similar type has also been discovered for
the thermodynamic entropy (2+1) dimensional black holes
recently \cite{carlip,teitelboim}.

The paper is arranged as follows. In Section 2, to keep
the paper self-contained, we discuss how edge states
arise for gauge theories defined on manifolds with spatial boundaries.
In Section 3, we discuss our model , namely
Maxwell-Chern-Simons ( MCS) theory defined on a disk coupled to a chiral scalar
field living on the edge of the disk.  This is a phenomenological model that
describes QHE (along with the edge currents present in a Hall system).  In
Section 4, we find the Hamiltonian for
the system by eliminating the unphysical modes using the Gauss law.
This Hamiltonian is quadratic and hence in principle,
one can find ground state
wavefunction \footnote{We thank R. Sorkin for emphasizing this point to us.
}.
In Section 5, we calculate the ground state using perturbation
theory and the entanglement entropy associated with
this ground
state.  We find that, in the above weak-coupling limit, this
entropy is approximately
proportional to the perimeter of the disk ( ``area law"). This result can also
be checked using the so-called
replica trick \cite{Edwards}, which we discuss in the appendix.

\noindent
\section{Edge Observables}

In this Section, we will see how edge observables occur for gauge theories on
manifolds with boundaries.  We will take the MCS theory on a disk as a typical
example \cite{mcs}
since we will be using this model also in the later Sections.

Here, we use the following conventions:
\begin{enumerate}
\item Greek and Latin indices take values 0,1,2 and 1,2 respectively.
\item The three-dimensional metric $\eta_{\mu \nu}$ is specified by
its nonvanishing
entries \\$\eta_{00}=-1,\;\; \eta_{11}=\eta_{22}=+1$ while the
three-dimensional Levi-Civita symbol is $\epsilon ^{\mu \nu \lambda}$
with $\epsilon ^{012}=+1$.
\item The spatial metric is given by the Kronecker $\delta_{ij}$ symbol
while the
two-dimensional spacetime and spatial
Levi-Civita symbols are $\epsilon^{\mu \nu}$ and
 $\epsilon^{ij} \equiv \epsilon^{0ij}$.
\end{enumerate}

We will also assume that the disk has a circular boundary and radius $R$.

The MCS Lagrangian is
\begin{eqnarray}
L & = & \int_{D } d^{2}x {\cal L} \nonumber \\
{\cal L} & = & -\frac{t}{4} \,
F_{\mu \nu}F^{\mu \nu}-\frac{\sigma _H }{2}\,
\epsilon ^{\mu \nu \lambda}A_{\mu}\partial _{\nu}A_{\lambda} \; .
 \label{1}
\end{eqnarray}
The Poisson brackets (PB's) for (\ref{1}) are
\begin{eqnarray}
\{ A_{i}(x),A_{j}(y)\} & = & \{ \Pi _{i}(x),\Pi _{j}(y)\} =0, \nonumber \\
\{ A_{i}(x),\Pi _{j}(y)\} & = & \delta _{ij}\delta ^{2}(x-y). \label{6}
\end{eqnarray}
Here and in what follows, all fields are to be evaluated at some fixed
time while the PB's are at equal times.  Thus $x^{0}=y^{0}$ in (\ref{6})\@.
As usual these PB's will be replaced by commutation relations (CR's) in the
quantized theory.
Also $A$ and $\Pi$ give the magnetic field $B$ and the components $F_{0i}:=
E_{i}$ of
the electric field by the formulae
\begin{eqnarray}
B & = & \epsilon _{ij} \partial _{i}A_{j}, \nonumber \\
E_{i} & = & \frac{1}{t}(\Pi _{i}+\frac{\sigma _H}{2}\epsilon _{ij}A_{j}).
\label{7}
\end{eqnarray}
The Hamiltonian and Gauss law for (\ref{1}) in quantum theory are
\begin{eqnarray}
&~& H  =  \int_{D} d^{2}x \: {\cal H},  \nonumber \\
&~&{\cal H}  =  \frac{1}{2t} [(\Pi _{i} +\frac{\sigma _H}{2}\epsilon
_{ij}A_{j})^{2} + t^{2}(\epsilon _{ij}
\partial _{i}A_{j})^{2}], \; \nonumber\\
&~& G(\chi) |\cdot \rangle = 0 \mbox{ for } \chi|_{\partial
D} = 0 \; , \label{2}
\end{eqnarray}
where
\be
G(\chi) = -\int_{D} d^{2}x\,\partial
_{i}\chi ^{(0)}[\Pi _{i}-\frac{\sigma _H}{2}\epsilon_{ij}A_{j}]
\label{3}
\ee
and $|\cdot \rangle$ is any physical state.

The reason for the boundary condition on $\chi$ is that it is only with this
condition that vanishing of $G(\chi) |\cdot \rangle$ is implied by the usual
Gauss law $\partial_i ( \Pi_i -\frac{ \sigma_H}{2} \epsilon_{ij} A_j) \approx
0$ (after a partial
integration).  Note that it is necessary to rewrite this classical Gauss law
in terms of $G(\chi)$ in quantum theory. This is because $ \Pi_i -
\frac{\sigma_H}{2} \epsilon_{ij} A_j$ are operator valued {\it distributions} in
quantum theory so that their derivatives have to be interpreted by smearing
them with the derivatives of suitable test functions \cite{mcs2}.
But it is only when the test functions satisfy
the above boundary conditions that the smeared Gauss law restricted to the
classical context vanishes by the classical Gauss law\cite{mcs2}.

The edge observables $Q (\Lambda )$ are obtained from $G$ by
changing the boundary conditions on $\chi$. They are
\be
Q(\Lambda ) = -\int_{D} d^{2}x\,\partial
_{i}\Lambda [\Pi _{i}-\frac{\sigma _H}{2}\epsilon_{ij}A_{j}]
\; , \; \Lambda |_{\partial D} = {\mbox{not necessarily 0 }} .
\label{4}
\ee
They are the generators of the affine Lie groups $\tilde{L}U(1)$ and
have the commutators
\be
[Q(\Lambda ),Q(\Lambda ')] = -i\sigma _H
\int _{D}d^{2}x\epsilon _{ij}\partial _i\Lambda \partial _j\Lambda '\;  = -i
\sigma_H \int_{\partial D} \Lambda d \Lambda'.
\label{5}
\ee
The action of $Q(\Lambda )$ on
$|\cdot \rangle$ depends only on the boundary value of
$\Lambda $ since the difference between $Q(\Lambda )$ and $Q(\Lambda ')$ when
$\Lambda$ and $\Lambda '$ coincide at the boundary is a constraint and hence
annihilates the physical states.  Furthermore, it commutes with
observables localized within $D$, its action on these observables being that of
a $G(\chi)$ \cite{mcs2}. Hence it can be regarded as localized at
the edge.

\noindent
\section{MCS Action with the Chiral Scalar Field}

The MCS action we have considered till now suffers however from an
``anomaly''.  Under the gauge transformations
\begin{eqnarray}
&& A_i \rightarrow A_i +\partial_i \Lambda ,\nonumber\\
&& E_i \rightarrow E_i ,\label{3.1}
\end{eqnarray}
for arbitrary $\Lambda$, the MCS action $S_{bulk}$ is not gauge invariant:
\be
S_{bulk} \rightarrow S_{bulk} -\frac{\sigma _H}{2}\int _{\partial {\cal M}}
d^{2}x\epsilon ^{\mu\nu}\partial _{\mu}\Lambda A_{\nu} \label{3.2}
\ee
(${\cal M}$ being $D\times {\bf R}$).
If we require that physics be gauge invariant, then we must modify the MCS
action $S_{bulk}$ by a surface term as follows:
\begin{eqnarray}
&&S_{tot} =S_{bulk}+\frac{\sigma _H}{2q}\int _{\partial {\cal M}}d^2 x\epsilon
^{\mu\nu}\partial _{\mu}\phi A_{\nu}-\frac{l^2}{8\pi}\int _{\partial {\cal
M}}d^2 x (D_{\mu}\phi )(D^{\mu}\phi ), \label{3.3}\\
&&D_{\mu}\phi =\partial _{\mu}\phi -qA_{\mu}. \label{3.4}
\end{eqnarray}
Here $l^2$ is a positive constant and $q$ is the charge by which the field
$\phi$ living at the boundary is gauged. Also  $d^2x = dt \, R d\theta$, $R$
being the radius of $\partial D$ and $\theta \in [0, 2\pi]$ its angular
variable.  Under the gauge transformations
(\ref{3.1}), $\phi$ transforms as
\be
\phi \rightarrow \phi + q \Lambda ,\label{3.5}
\ee
so that $\Phi =e^{i\phi}$ transforms like a complex scalar field with charge
$q$.  [Since $\Phi$ is the true charged excitation, we also
have the identification $\phi \approx \phi +2\pi$.]

With this transformation law for $\phi$, we see that
\be
S_{tot}\rightarrow S_{tot} \label{3.6} .
\ee

So far, the coefficient $l^2$ in (\ref{3.3}) is arbitrary.  However, if this
action is to describe QHE \cite{callan,dhe}, then the currents due to the edge
scalar field $\phi$ are required to be ``chiral''.  In this case, it can be
shown \cite{dhe} that the coefficient $l^2$ gets uniquely fixed according to
\cite{callan}
\be
 l^2 = 2\pi\frac{\sigma_H}{q^2}
\label{3.6a}
\ee

  With $l^2$ fixed in this way, we can impose the
``chirality'' constraint that the edge fields are left-moving, namely that
\be
D_{0}\phi - D_{\theta}\phi =0. \label{3.7}
\ee
Thus the total action that correctly describes QHE is
\be
S_{tot} =S_{bulk} + \frac{\sigma _H}{2q}\int _{\partial {\cal M}}d^2 x\epsilon
^{\mu\nu}\partial _{\mu}\phi A_{\nu}-\frac{\sigma _H}{4q^2}\int _{\partial {\cal
M}}d^2 x (D_{\mu}\phi )(D^{\mu}\phi ). \label{3.8}
\ee

\noindent
\section{The Hamiltonian and Quantization}
In this Section, we will find the Hamiltonian for the action (\ref{3.8}) and
then set up the formalism for the quantization of this interacting system.

From the action (\ref{3.8}), we find the momentum conjugate to $A_i$ to be
\be
\Pi _i = tE_i -\frac{\sigma _H}{2}\epsilon _{ij}A_j \label{4.1}
\ee
(where $E_i := F_{0i}$) and the momentum conjugate to $\phi$ to be
\be
\Pi _{\phi} = \frac{\sigma _H}{2q^2}D_0 \phi + \frac{\sigma _H}{2q}A_{\theta}
.\label{4.2}
\ee
The Hamiltonian for this system is
\begin{eqnarray}
&&H_{tot}=H_{bulk} +H_{edge}    \nonumber\\
&&H_{bulk}=  \frac{1}{2t} \int _{D}d^2 x[(\Pi _i +\frac{\sigma _H}{2} \epsilon
_{ij}A_j )^2 +t^2 (\epsilon _{ij}\partial _i A_j )^2]\nonumber\\
&& H_{edge}= \frac{1}{2}\int _{\partial D}dx[\frac{2q^2}{\sigma _H}(\Pi_{\phi}
-\frac{\sigma _H}{2q}A_{\theta})^2 +\frac{\sigma _H}{2q^2}(\phi
'-qA_{\theta})^2]\label{4.3}
\end{eqnarray}
and the Gauss law constraint is
\be
{\cal G}(\chi ):= -\int _{D}d^2 x\partial _i \chi (\Pi _i -\frac{\sigma
_H}{2}\epsilon _{ij}A_j )-q\int _{\partial D} R d\theta
\, \chi (\Pi _{\phi} +\frac{\sigma
_H}{2q^2}\phi ') \approx 0,\label{4.4}
\ee
$R$ being the radius of the disk and $\theta$(mod $2\pi$) the angle coordinate
for $\partial D$.
To quantize the system described by (\ref{4.3}) and (\ref{4.4}), we now need to
choose mode expansions for the fields $A_i ,\Pi _i, \phi$ and $\Pi _{\phi}$.
As usual, the coefficients of these modes will play the role of creation and
annihilation operators in the quantized theory.

In an earlier work \cite{mcs,mcs2},
we have quantized the bulk theory alone (namely the MCS action {\em without}
the chiral scalar field).

There we showed that among the various permissible
mode expansions parametrised by a real parameter $\lambda$, only the one
characterized by $\lambda =0$ allows the existence of edge observables.  This
is because each $\lambda$ specifies a domain ${\cal D}_{\lambda}$ for the space
of one-forms ($A_i$ or $\Pi _i$) and the edge observables do not leave this
domain invariant unless $\lambda =0$.  Since edge observables are crucial in
our work here, we will use only the mode expansions specified by the choice
$\lambda =0$.

The corresponding mode expansions for $A$, $\Pi$ and $E$ are \cite{mcs}
\begin{eqnarray}
&&A= (a_{nm}\frac{\Psi _{nm}}{\sqrt{t}}+a_{nm}^{(*)}\frac{*\Psi _{nm}}{\sqrt{t}}
+ \alpha _n \frac{h_n}{\sqrt{\sigma _H}}) +\;\; h.c.,\label{4.5}\\
&&\Pi = ({\pi}_{nm}\sqrt{t}\Psi _{nm} +{\pi}_{nm}^{(*)}\sqrt{t}*
\Psi _{nm}+p_n \sqrt{\sigma _H}h_n )+\;\; h.c.,\label{4.6}\\
&&E = (e _{nm}\frac{\Psi _{nm}}{\sqrt{t}}+e_{nm}^{(*)}\frac{*\Psi
_{nm}}{\sqrt{t}})+\frac{\sqrt{\sigma _H}}{t}(p_n +\frac{i}{2}\alpha _n )h_n+
h.c.\;, \label{4.7} \\
&&e_{nm}:= \pi_{nm} - \frac{\sigma_H}{2t}a_{nm}^{(*)}, \nonumber \\
&&e_{nm}^{(*)}= \pi_{nm}^{(*)}+ \frac{\sigma_H}{2t}a_{nm}, \nonumber 
\end{eqnarray}
summation over repeated indices [ $n$ over positive and $m$ over 
non-negative integers] being understood.
Here, we have used (\ref{4.1}) and also
 the notation of forms while $*$ refers to the Hodge dual
\cite{8}. Further we have scaled the modes of \cite{mcs}
 for later convenience
and dropped
the superscript $(1)$ employed in \cite{mcs} to emphasize one-forms.
The modes used above are defined as follows ( see \cite{mcs} for details):
\begin{eqnarray}
&&\Psi _{nm} :=N_{nm}*d(J_n (\omega _{nm}r)e^{in\theta}), \nonumber\\
&&h_n :=\frac{1}{\sqrt{2\pi n}R^n}d(r^n e^{in\theta}). \label{4.71}
\end{eqnarray}
The $d$ refers to the exterior derivative, the $J_n$'s refer as
usual to the cylindrical Bessel functions \cite{10} while the $\omega _{nm}$'s
are such that $J_n (\omega _{nm}R)=0$.
Also the $N_{nm}$'s are normalization
constants chosen such that $\int _{D}\bar{\Psi } _{nm}*\Psi _{nm}=-1$, bar
denoting complex conjugation.

In (\ref{4.5}-\ref{4.7}), it is understood that
$n$ is summed over all non-negative integers for the
first two terms and
all the positive integers for the third term while $m$ is summed over all
positive integers.

From the PB relations (\ref{6}) and using also {\ref{7}), we get the
following non-vanishing CR's:
\begin{eqnarray}
&& [a_{nm},{\pi}^{\dagger}_{n'm'}]=[a_{nm}^{\dagger},{\pi}_{n'm'}]
=[a_{nm},e ^{\dagger}_{n'm'}]=[a^{\dagger}_{nm},e _{n'm'}] =i\delta _{nn'}
\delta _{mm'} ,\nonumber\\
&& [a_{nm}^{(*)},{\pi}^{(*)\dagger}_{n'm'}]=[a_{nm}^{(*)\dagger},
{\pi}_{n'm'}^{(*)}]
=[a_{nm}^{(*)},e ^{(*)\dagger}_{n'm'}]=[a^{(*)\dagger}_{nm},e _{n'm'}^{(*)}
] =i\delta _{nn'}
\delta _{mm'} ,\nonumber\\
&& [e_{nm},e_{n'm'}^{(*)\dagger}]=[e_{nm}^{\dagger}, e
_{n'm'}^{(*)}]=i\frac{\sigma _H}{t}\delta _{nn'}\delta _{mm'}, \nonumber\\
&&[\alpha _{n}, p_{n'}^{\dagger}]=[\alpha _n ^{\dagger},p_{n'}] =i\delta
_{nn'}. \label{4.75}
\end{eqnarray}

We also mode expand the edge fields $\phi$ and $\Pi _{\phi}$ as follows:
\begin{eqnarray}
&&\phi = (\phi _n \frac{q}{\sqrt{2\pi n\sigma _H}}e^{in\theta})+\;\; h.c.,
\label{4.8}\\
&&\Pi _{\phi}= (\pi _n\frac{\sqrt{2\pi n\sigma _H}}{2\pi qR}e^{in\theta})
+\;\; h.c..\label{4.9}
\end{eqnarray}
We have suppressed the winding modes in writing (\ref{4.8},\ref{4.9})
as they are not
important for this paper. Also $n$ here is summed only
over non-negative integers.  Though (\ref{4.8}) looks
singular for $n=0$, it is all right for our
purposes since
what occurs in the Hamiltonian and the Gauss law is $\phi '$ (and not $\phi$)
and this
itself admits a well-defined mode expansion even for $n=0$.  Once again, the
only non-zero CR's are
\be
[\phi _{n},\pi^{\dagger}_{n'}] =i\delta _{nn'} .\label{4.95}
\ee

The Gauss law (\ref{4.4}) in terms of these modes takes the form
\be
(e_{nm}^{(*)}-\frac{\sigma _H}{t}a_{nm})| \cdot \rangle =
(e_{nm}^{(*)}-\frac{\sigma _H}{t}a_{nm})^{\dagger}
| \cdot \rangle =0 \label{4.10}
\ee
and
\be
\left\{(p_n -\frac{i}{2}\alpha _n )+(\pi _n +\frac{i}{2}\phi _n ) \right\}
|\cdot \rangle =
\left\{(p_n ^{\dagger}+\frac{i}{2}\alpha _n ^{\dagger}) +(\pi _n ^{\dagger}-
\frac{i}{2}\phi _n ^{\dagger}) \right\}|\cdot \rangle = 0 ,\label{4.11}
\ee
$|\cdot \rangle$ being any physical state. Here (\ref{4.10}) is the condition
on the
modes $^*\Psi_{nm}$ while (\ref{4.11}) arises as the constraint on the modes
 $h_n, {\bar{h}_n}$.

We can use (\ref{4.10}) to eliminate $e^{*}_{nm}$
in favour of $a_{nm}$.
In that case the Hamiltonian (\ref{4.3}) in terms of these modes becomes
\begin{eqnarray}
H_{tot}&=&H_{bulk}+H_{edge} \nonumber\\
H_{bulk}&=& \frac{1}{2} \left\{ e_{nm}^{\dagger}e_{nm}
+(\omega _{nm}^2+(\frac{\sigma
_H}{t})^2 )a^{\dagger}_{nm}a_{nm}\right\}+\frac{\sigma _H}{2t}(c_n^{\dagger}c_n + c_n
c_n^{\dagger} ) \nonumber\\
H_{edge} &=& \frac{n}{R}[(\pi _n +\frac{i}{2}\phi _n -i\alpha _n
+R\sqrt{\frac{2\pi \sigma _H}{nt}}\omega _{nm}N_{nm}J_n '(\omega _{nm}R)a
_{nm})^{\dagger}\nonumber\\
&&(\pi _n +\frac{i}{2}\phi _n -i\alpha _n
+R\sqrt{\frac{2\pi \sigma _H}{nt}}\omega _{nm'}N_{nm'}J_n '(\omega _{nm'}R)a
_{nm'}) + (\pi _n -\frac{i}{2}\phi _n )^{\dagger}(\pi _n -\frac{i}{2}\phi
_n )]\label{4.12}
\end{eqnarray}
where
\be
c_n \equiv (p_n +\frac{i}{2} \alpha_n)^\dagger,
\label{c}
\ee
and
\be
J'_n(x) \equiv \frac{d}{d x} J_n(x).
\label{d}
\ee
If we are in one chiral sector of the edge theory (the physically relevant
sector when dealing with QHE), then we can also impose the condition
\be
(\pi _n -\frac{i}{2}\phi _n )|\cdot \rangle =0\label{4.13}
\ee
on physical states $|\cdot \rangle$ (such a condition arising from the mode expansion of the
chirality constraint (\ref{3.7})).

We can now also use the Gauss law (\ref{4.11}) to eliminate $(\pi _{n}
+\frac{i}{2}\phi _n )$ in terms of $(p_n -\frac{i}{2}\alpha _n )$.  The edge
Hamiltonian then reduces (in the above chiral sector) to
\begin{eqnarray}
H_{edge}&=& \frac{n}{R}[(p_n +\frac{i}{2}\alpha _n -R\sqrt{\frac{2\pi\sigma
_H}{nt}}\omega _{nm}N_{nm}J_n '(\omega _{nm}R)a_{nm})^{\dagger}\nonumber\\
&&(p_n +
\frac{i}{2}\alpha _n -R\sqrt{\frac{2\pi\sigma _H}{nt}}\omega _{nm'}N_{nm'}
J_n '(\omega _{nm'}R)a_{nm'})].
\label{4.14}
\end{eqnarray}
Therefore the task of quantization reduces to the diagonalization of the system
described by the following Hamiltonian :
\begin{eqnarray}
H &=& \frac{1}{2}\left\{e_{nm}^{\dagger}e_{nm} +(\omega _{nm}^2
+(\frac{\sigma _H}{t})^2
)a_{nm}^{\dagger}a_{nm} \right\} +\frac{\sigma _H}{t}c_n ^{\dagger}c_n
+\nonumber\\
&&\frac{n}{R}(c_n^{\dagger} -R\sqrt{\frac{2\pi\sigma _H}{nt}}\omega _{nm}N_{nm}
J_n '(\omega _{nm}R)a_{nm} )(c_n -R\sqrt{\frac{2\pi\sigma
_H}{nt}}\omega _{nm'}N_{nm'} J_n '(\omega _{nm'}R)a_{nm'}^{\dagger} ).
\label{4.15}
\end{eqnarray}
The Hamiltonian in (\ref{4.15}) has also been normal ordered, the vacuum
$|0\rangle$ being defined by $e_{nm} |0\rangle$ = $a_{nm}|0\rangle$ = $c_n |
0\rangle=0$. Note that $H$ preserves the constraints (\ref{4.10}),(\ref{4.11})
and (\ref{4.13}).
\noindent
\section{The Ground State and its Entanglement Entropy}

We are not interested here in finding the entire spectrum of the Hamiltonian
(\ref{4.15}).  Our objective is in finding the ground state of the system.  As
is clear even from (\ref{4.15}), this system has correlations between the edge
degrees of freedom  $a_n$ and
bulk degrees of freedom $e_{nm},\; a_{nm}$.  The density
matrix obtained by tracing out the edge states will
therefore be impure and will have a non-zero entropy associated to it.

We will now find the ground state of this interacting system by using
perturbation theory with $q$ being the perturbing parameter.  Before starting
the calculations, it is important to note that in physical situations in
QHE, the coefficient $\sigma _H$ can be written as $kq^2$ where $k$ is a number
of order unity.  We will therefore replace $\sigma _H$ in (\ref{4.15}) by
$kq^2$ to make the order in $q$ explicit:
\begin{eqnarray}
H &=& \frac{1}{2} \left\{e_{nm}^{\dagger} e_{nm} +(\omega _{nm}^2
+(\frac{kq^2}{t})^2
)a_{nm}^{\dagger}a_{nm} \right\} +\frac{kq^2}{t}c_n^{\dagger}c_n +\nonumber\\
&& \frac{n}{R}(c_n^{\dagger} -qR\sqrt{\frac{2\pi k}{nt}}\omega _{nm}N_{nm}
J_n '(\omega _{nm}R)a_{nm} )(c_n -qR\sqrt{\frac{2\pi k}{nt}
}\omega _{nm'}N_{nm'} J_n '(\omega _{nm'}R)a_{nm'}^{\dagger} ).\label{5.1}
\end{eqnarray}
Keeping terms only to order $q^2$ and defining $\hat{e} _{nm}=e
_{nm}/\sqrt{\omega _{nm}}$ and $\hat{a}_{nm}=a_{nm}\sqrt{\omega _{nm}}$, the
above
Hamiltonian can be rewritten as
\begin{eqnarray}
H&=&\frac{1}{2}\omega _{nm}[\hat{e}_{nm}^{\dagger}\hat{e}_{nm}+
\hat{a}_{nm}^{\dagger}
\hat{a}_{nm}]+ (\frac{n}{R}+\frac{kq^2}{t})c_n ^{\dagger}c_n -
q\sqrt{\frac{2\pi nk\omega _{nm}}{t}}N_{nm} J_n '(\omega _{nm}R)(c
_n\hat{a}_{nm} +c_n^{\dagger} \hat{a} _{nm}^{\dagger})+ \nonumber\\
&&q^2 \frac{2\pi
kR}{t}\sqrt{\omega _{nm}\omega _{nm'}}N_{nm}N_{nm'}J_n '(\omega _{nm}R)J_n
'(\omega _{nm'}R)\hat{a}_{nm}\hat{a} _{nm'}^{\dagger} \label{5.2}
\end{eqnarray}
We now define annihilation operators ${\cal A}_{nm}$ and ${\cal B}_{nm}$ as
follows:
\begin{eqnarray}
&&{\cal A}_{nm}:=\frac{1}{\sqrt{2}}(\hat{e} _{nm}-i\hat{a}_{nm}) \nonumber \\
&&{\cal B}_{nm}:=\frac{1}{\sqrt{2}}(\hat{e}_{nm}^{\dagger}-
i\hat{a}_{nm}^{\dagger}). \label{5.3}
\end{eqnarray}

Their non-vanishing commutators are contained in
\bea
&&[ {\cal A}_{nm},{\cal A}_{n'm'}^{\dagger}] = \delta_{nn'} \delta_{mm'},
\nonumber \\
&&[{\cal B}_{nm},{\cal B}_{n'm'}^{\dagger}] = \delta_{nn'} \delta_{mm'}
\label{5.3a}
\eea

In terms of these variables the Hamiltonian is
\begin{eqnarray}
H&=&\frac{1}{2}
\omega _{nm}[{\cal A}_{nm}^{\dagger}{\cal A}_{nm}+{\cal B}_{nm}^{\dagger}
{\cal B}_{nm}]+\frac{n}{R}c_n ^{\dagger}c_n -\nonumber\\
&& \frac{q}{i}\sqrt{\frac{\pi nk\omega _{nm}}{t}}N_{nm} J_n '(\omega
_{nm}R)(c_n ({\cal B}_{nm}^{\dagger}-{\cal A}_{nm})+c_n^{\dagger}
({\cal
A}_{nm}^{\dagger} -{\cal B}_{nm})) +\nonumber\\
&& q^2 [\frac{k}{t}c_n ^{\dagger}c_n +\frac{\pi kR}{t}\sqrt{\omega
_{nm}\omega _{nm'}}N_{nm}N_{nm'}J_n '(\omega _{nm}R)J_n '(\omega _{nm'}R)({\cal
A}_{nm}^{\dagger}{\cal A}_{nm'}+{\cal B}_{nm'}^{\dagger}{\cal B}_{nm}
\nonumber\\
-&&{\cal A}_{nm}^{\dagger}{\cal B}_{nm'}^{\dagger}
-{\cal B}_{nm}{\cal A}_{nm'})]
\label{5.4}
\end{eqnarray}
upto additive constants caused by ordering of operators.

We can now find the ground state $|\Psi \rangle _{gnd}$ to order $q^2$. For 
convenience  let us fix the angular momentum $n$. The true ground state will
be a product of wavefunctions with different $n$'s. Therefore, 
\be
| \Psi \rangle_{gnd} = \prod_n |\Psi, n  \rangle_{gnd}, 
\label{prod}
\ee
where,
\begin{eqnarray}
|\Psi ,n \rangle _{gnd}&=& |0\rangle 
[1-\frac{q^2}{2}\sum _{m}\frac{\pi nk\omega_{nm}
N^2 _{nm}(J_n '(\omega_{nm}R))^2}{t(\frac{n}{R}+\frac{\omega_{nm}}{2})^2}]
+\nonumber\\
&& \sum _{m}|1\rangle_{c_n}|1\rangle _{{\cal A}_{nm}}(iq
\sqrt{\frac{\pi nk\omega_{nm}}{t}}\frac{N_{nm}J_n '
(\omega _{nm}R)}{\frac{n}{R}+\frac{\omega _{nm}}{2}}) +\nonumber\\
&&q^2 \sum _{m,m'}|1\rangle _{{\cal A}_{nm}}|1\rangle _{{\cal
B}_{nm'}}\frac{\pi k}{t}\frac{\sqrt{\omega_{nm}\omega _{nm'}}N_{nm}N_{nm'}J_n
'(\omega _{nm}R)J_n '(\omega _{nm'}R)}{(\frac{\omega _{nm} +\omega
_{nm'}}{2})}(\frac{\omega _{nm}R}{\frac{2n}{R}+\omega _{nm}})-\nonumber\\
&&q^2 \sum _{m,m'}|2\rangle_{c_n}|1\rangle _{{\cal A}_{nm}}|1\rangle
_{{\cal A}_{nm'}}(\frac{\pi nk}{t})\frac{\sqrt{2\omega _{nm}\omega
_{nm'}}N_{nm}N_{nm'}J_n '(\omega _{nm}R)J_n '(\omega
_{nm'}R)}{(\frac{n}{R}+\frac{\omega _{nm}}{2})(\frac{2n}{R}+\frac{\omega _{nm}
+\omega_{nm'}}{2})}.\label{5.5}
\end{eqnarray}
Here $|0\rangle$ refers to the unperturbed ground state :
\be
{\cal A}_{nm}|0\rangle = {\cal B}_{nm} |0\rangle = c_n |0\rangle =0.
\label{5.5a}
\ee
Also the state
$|1\rangle _{{\cal A}_{nm}}$ refers to the first excited state created by the
creation operator ${\cal A}_{nm}^{\dagger}$, a similar notation being used
for the other states.

From this ground state, we can also now define the corresponding density matrix
${\hat{\rho}}$ and then obtain
the reduced density matrix by tracing over the bulk states.
[Actually, we may want to trace over the edge states if the objective is
to calculate the entropy due to a lack of knowledge of the edge states.  But
the answer for the entropy itself is the same whether we trace over one or the
other.]      To order $q^2$, we have for this reduced density matrix,
\be
\hat{\rho} _{red}:=Tr_{(bulk)} (\hat{\rho})=|0\rangle _{a_n\;a_n}\langle 0|
(1-{\bf p}_n )+|1\rangle _{a_n\; a_n}\langle 1| {\bf p}_n \label{5.6}
\ee
where
\be
{\bf p}_n =q^2 \sum _{m}[(\frac{\pi nk}{t})\frac{\omega _{nm}N^2 _{nm}J_n
^{'2}(\omega _{nm}R)}{(\frac{n}{R}+\frac{\omega _{nm}}{2})^2}]. \label{5.7}
\ee
 The dependence on $R$ factors out in the expression for
${\bf p}_n$ as can be seen by rewriting it as follows:
\begin{eqnarray}
{\bf p}_n &=& q^2 (\pi nk)\frac{R}{t}\sum _m \frac{(\omega _{nm}R)N^2 _{nm} J_n
^{'2}(\omega _{nm}R)}{(n+\frac{\omega _{nm}R}{2})^2}\nonumber\\
&=& q^2 (\pi n k ) \frac{R}{t} \sum_m 4 \frac{ \chi_{nm} N^2_{nm}
J'^2_n(\chi_{nm})}{ (2n +\chi_{nm})^2} \nonumber \\
&\rightarrow & q^2 \frac{4\pi nkR}{t}\sum
_m \frac{1}{(2n+\chi _{nm})^2 \chi _{nm}} \mbox{   for large $\chi _{nm}:=\omega
_{nm}R$,}
\label{5.8}
\end{eqnarray}
$\chi_{nm}$ being zeroes of $J_n$. This shows that ${\bf p}_n$ is
proportional to $R$. Note that ${\bf p}_n$ is finite.
This is because for large $m$,
\be
| \chi_{nm}| \sim \frac{\pi}{2} ( 2m + n - \frac{1}{2}),
\qquad \qquad  N_{nm} J'_{n} ( \chi_{nm}) \sim O (\frac{1}{\chi_{nm}})
\label{5.8a}
\ee
so that
\be
\frac{ \chi_{nm} N^2_{nm} J'^2_{n}(\chi_{nm})}{(2n+ \chi_{nm} )^2} =
O(\frac{1}{m^3})
\label{5.9}
\ee
Here the above behavior of $\chi_{nm}$, and the behavior  of
 $J_n(\chi_{nm})$ and $J_n'(\chi_{nm})$ for large
$|\chi_{nm}|$ are known from the literature \cite{abramowitz} to be

\bea
J_n(\chi_{nm}) \sim \sqrt{\frac{2}{\pi \chi_{nm}}} \cos ( \chi_{nm} -
\frac{n\pi}{2} - \frac{\pi}{4}) =  O(\frac{1}{\sqrt{\chi_{nm}}}), \nonumber \\
J'_n(\chi_{nm}) \sim - \sqrt{\frac{2}{\pi \chi_{nm}}} \sin ( \chi_{nm} -
\frac{n\pi}{2} - \frac{\pi}{4}) =  O(\frac{1}{\sqrt{\chi_{nm}}})
\label{5.10}
\eea
while that of $N_{nm}$ is obtained from the normalization condition
\be
\int \int _D N^2_{nm} \left[\frac{\chi_{nm}}{R}^2 J'^2_n (\frac{ \chi_{nm}r}{R}) +
\frac{n^2}{r^2} J^2_n
(\frac{\chi_{nm}r}{R})  \right] r dr d\theta = 1
\label{5.11}
\ee
on $\Psi_{nm}$.
Hence,
\be
\frac{1}{N^2_{nm}} = 2\pi \int_0^{\chi_{nm}} \left[
\left(\frac{dJ_n (u)}{du}\right)^2+ (\frac{n}{u}
J_n(u))^2 \right] \;u\, du ,
\label{5.12}
\ee
where $ u= \frac{\chi_{nm}r}{R}$.
It shows that $N^2_{nm}$ depends only on $\chi_{nm}$ and gives
\be
\frac{\partial({\frac{1}{N^2_{nm}}})}{\partial \chi_{nm}} =  \mbox{ constant
 for large $|\chi_{nm}|$}
\label{5.13}
\ee
Hence
\be
N_{nm} =O(\frac{1}{\sqrt{\chi_{nm}}}) \qquad \mbox{as $|\chi_{nm}| \ra
\infty$}
\label{5.14}
\ee
Putting these estimates together we get (\ref{5.9})

We can now infer that ${\bf p}_n$ being finite, the entropy of the $n$th mode
\be
S_n :=-{\bf p}_n \log {\bf p}_n-(1-{\bf p}_n )\log (1-{\bf p}_n ) \label{5.A}
\ee
is also finite for generic ${\bf p}_n$.

[However, the total entropy defined as $S:= \sum
_n S_n$ can diverge unless one imposes a cut-off (for example, on the maximum
allowed $n$).]

\noindent
\section{Concluding Remarks}

In this concluding section, we draw attention to the approximate
``area law'' obeyed by the
above entanglement entropy.  From the expression (\ref{5.A}) for the entropy
$S_n$, we see that in the limit of $q$ being small, we can approximate it by
just the first term.  Namely,
\be
S_n \approx {\bf p}_n-{\bf p}_n \log {\bf p}_n \qquad \mbox{ for $q$ small} \label{6.1}
\ee
Since we know from (\ref{5.8}) that ${\bf p}_n$ itself scales as $R$,
we see that
$S_n$ also scales as $R$ (apart from the logarithmic term).  For a disk, this
is therefore a statement that the entropy approximately
scales like the perimeter of the disk
(or the area of the boundary of the disk).

It is interesting that such an ``area law'' has emerged for the entanglement
entropy of edge states.  Before we draw general conclusions, however,
we should
check whether such a law is a mere coincidence for the 2+1 theory describing
Hall effect or whether it holds more generally.  One possibility is to find the
entanglement entropy for a similar theory in 3+1 dimensions, but without a
Chern-Simons term.  If the ``area law'' does hold in such more general
situations, it seems reasonable to conclude that the black hole entropy
is at least partially the creation of the entanglement entropy arising out of
a lack of
knowledge of the edge states describing excitations of the black hole horizon.

\noindent
\section*{Acknowledgements}
It is a great pleasure to thank Rafael Sorkin for valuable comments.
We are also grateful to  G. Jungman,
B. Sathiapalan, S. Vaidya and J. Varghese for discussions.
This work was supported
by the US Department of Energy under
 contract numbers DE-FG02-85ER40231 and  DE-FG05-86ER40272
and by a Syracuse University Graduate Fellowship awarded to A.M.

\newpage
\appendix
\section*{The Replica Trick in Action}

In this section we show that the results given above also follows
from another method, namely the replica trick \cite{Edwards}. 
Our Hamiltonian, given by (\ref{4.15}), is quadratic in the various field
modes and their conjugate momenta. The ground state wavefunctional
for this system
will then be a generalized Gaussian. To simplify the problem and to
highlight the
important aspects, we will restrict ourselves to one tower of
radial modes $a_{nm}$ with any fixed $m$, which
we will call $a_n$, and
keep all the edge modes $\alpha_n$. 
The measure defining the scalar product is 
$\prod_n da_n da^\dagger_n d \alpha_n d{\alpha}^\dagger_n$. 
The calculation is performed using the Schr\"{o}dinger representation
where we have the substitutions
\bea
e_{n} \ra -i \frac{\partial}{\partial a^\dagger _{n}} \qquad
e^\dagger _{n} \ra -i \frac{\partial}{\partial a_{n}} \label{a.1a} \\
p_n \ra -i\frac{\partial}{\partial \alpha^\dagger _n}\;, \qquad
p^\dagger _n \ra -i\frac{\partial}{\partial \alpha_n} \label {a.1b}.
\eea
The ground state wavefunctional is found by
using the following Gaussian ansatz which is consistent with
angular momentum conservation:
\be
\bbox{ \Psi }(\{a_n\},\{a^\dagger_n\};\{\alpha_n\},\{\alpha^\dagger_n\}) =  
\prod_{n}{\cal N}_n e^
{-( A_n \alpha_n \alpha^\dagger_n+ B_n a_n a^\dagger_n + C_n a_n^\dagger
\alpha_n
+ D_n
 \alpha_n^\dagger a_n)}.
\label{a.2}
\ee
Here $\prod_{n}{\cal N}_n$ is the normalization factor for the wavefunction and
$A_n,B_n,C_n,D_n$ are real constants to be determined. It is also understood
that $\dagger$ denotes complex conjugation here.

The Schr\"{o}dinger equation is
\be
H \bbox{\Psi}(\{a_n\}, \{\alpha_n\} ) = E\bbox{\Psi}(\{a_n\}, \{\alpha_n\})
\label{a.3}
\ee
where our Hamiltonian
is given by (\ref{4.15}).
The constants $A_n,B_n,C_n$ and $D_n$ in the 
wavefunctional $\bbox{\Psi}$ can be determined from (\ref{a.3}) and 
requiring the wavefunction to be normalizable. $A_n,C_n$ 
and $D_n$ turn out to be 
\bea
A_n &=& \frac{1}{2}\; , \nonumber \\
C_n &=&-\sqrt{\frac{2\pi \sigma_H}{n t}} \frac{2 i n
\omega_n N_{nm} J'_n(\chi_{nm})}{(2E_n+\frac{\sigma_H}{t}+ \frac{n}{R})} \;,
\nonumber \\
D_n&=&0 \; ,
\label{a.4}
\eea
while $B_n$ is determined by a complicated cubic equation.
The solutions to this equation are rather cumbersome  and are
not illuminating.
However, in the weak coupling ($\frac{\sigma_H}{t} \ra 0$)
and large radius limit ( $R \ra \infty$),
one gets the unperturbed value,
\be
B_n = \omega_n.
\label{a.5}
\ee

The ground state energy is given by $E=\sum_n E_n$ where 
\be
E_n = (\frac{\sigma_H}{t} + \frac{n}{R})A_n +\frac{1}{2} B_n,
\label{energy}
\ee

Hereafter we will write our wavefunctional as 
\be
\bbox{ \Psi }(\{a_n\},\{a^\dagger_n\};\{\alpha_n\},\{\alpha^\dagger_n\}) =  
\prod_{n}{\cal N}_n e^{-( A_n \alpha_n \alpha^\dagger_n+ 
B_n a_n a^\dagger_n + C_n a_n^\dagger \alpha_n)}.
\label{a.6}
\ee
as $D_n=0$.

Given the above
form for the wavefunctional, one can readily form the ground state
density matrix $\bbox{\rho}$ :
\bea
\bbox{\rho}(\{a_n\};\{\alpha_n\}|\{a'_n\};\{\alpha'_n\}
)&=& \bbox{ \Psi }(\{a_n\};\{\alpha_n\})
\bbox{\Psi}^*
(\{a'_n\};\{\alpha'_n\}) \nonumber \\
&& = \prod_{n} |{\cal N}_n|^2 e^
{-\{ A_n (\alpha_n \alpha^\dagger_n+\alpha'_n \alpha'^{\dagger}_n)+ 
B_n (a_n a^\dagger_n + 
a'_n a'^{\dagger}_n)
 + C_n (a^\dagger_n \alpha_n + \alpha'^{\dagger}_n a'_n)\}} \nonumber \\
&& \equiv  \prod_n \bbox{\rho}^{(n)}(a_n,\alpha_n|a'_n,\alpha'_n).
\label{a.7}
\eea

The reduced density matrix is found by tracing over the set of variables
which is not being observed , say ${a_n}$'s. Therefore, the 
reduced density
matrix in this case would be given by
\be
\bbox{\rho}_{red.} ( \{\alpha_n\};\{\alpha'_n\})= \prod_{n}
\int da_n d{\bar{a}_n}
\bbox{\rho}^{(n)}(a_n;\alpha_n|a_n;\alpha'_n).
\label{a.8}
\ee
A  straightforward Gaussian integration over the complex variables ${a_n}$
leads
to
\bea
\bbox{\rho}_{red.}(\{\alpha_n\};\{\alpha'_n\}) &=&\prod_n |{\cal N}'_n|^2 e ^{-
A_n\{\alpha_n \alpha^\dagger_n+ \alpha'_n \alpha'^{\dagger}_n \}+ 
\gamma_n
\alpha'^{\dagger}_n \alpha_n }, \nonumber \\
&\equiv & \bbox{\rho}_{red}^{(n)}(\alpha_n; \alpha'_n)
\label{a.9}
\eea
where ${\cal N'}_n = {\cal N}_n\sqrt{\frac{\pi}{2B_n}}$ and $\gamma_n = 
\frac{|C_n|^2}{2B_n}$. The normalization factor ${\cal N'}$ can
be computed easily by recalling that for any density matrix $\bbox{\rho}$,
we require
$\tr \bbox{\rho} = 1$. This gives
\be
|{\cal N'}_n|^2 = \frac{1}{\pi} (2 A_n - \gamma_n )
\label{a.10}
\ee

The entanglement entropy $S= - \tr \bbox{\rho}_{red} \ln \bbox{\rho}_{red}$
can be computed from the reduced density matrix
using the so-called replica trick \cite{Edwards} according to which
\be
S =  - \frac{d}{dN}( \tr (\bbox{\rho}_{red}^N
))|_{N=1},
\label{a.11}
\ee
$\bbox{\rho}_{red}$ here being the operator with the kernel (\ref{a.8}).

Now, one can find for real and positive $A$ the following equation
\bea
&&\bbox{\rho}_{red}^N (\{\alpha_n\}; \{\alpha'_n\})\equiv  \prod_n
\int \prod_{i=1}^{N-1}dz_{i,n} d\bar{z}_{i,n} \bbox{\rho}_{red}^{(n)}
(\alpha_n;z_{1,n})
\bbox{\rho}_{red}^{(n)}(z_{1,n};z_{2,n})\cdots \bbox{\rho}_{red}^{(n)}
(z_{N-1,n};\alpha'_n) \nonumber
\\
&=&\prod_n |{\cal N}'_n|^{2N} e^{-A_n \alpha_n \alpha_n^\dagger}
\int dz_{1,n}d\bar{z}_{1,n}e^{-2A_n z_{1,n}\bar{z}_{1,n}+\gamma_n
\bar{z}_{1,n} \alpha_n} \int dz_{2,n}d\bar{z}_{2,n}e^{2A_n z_{2,n}
\bar{z}_{2,n}+\gamma_n\bar{z}_{2,n} z_{1,n}}\int dz_{3,n}d\bar{z}_{3,n}
 \times 
\cdots \nonumber \\
&& \qquad \cdots \times \int dz_{N-1,n}
d\bar{z}_{N-1,n} e^{-2A_n z_{N-1,n}\bar{z}_{N-1,n}+\gamma_n\alpha'^\dagger_n 
z_{N-1,n}}e^{-A_n \alpha'_n \alpha'^{\dagger}_n} \nonumber \\
&=&\prod_n |{\cal N}'_n|^{2N} e^{-A_n (\alpha_n \alpha_n^\dagger+\alpha'_n
\alpha'^\dagger_n)} \int dz_{N-1,n}d\bar{z}_{N-1,n} e^{-2A_n z_{N-1,n}
\bar{z}_{N-1,n} + \gamma_n(\alpha'^\dagger_n z_{N-1,n}+ \bar{z}_{N-1,n}
z_{N-2,n})}\times \cdots \nonumber \\
\nonumber \\
&& \qquad\cdots \times \int dz_{1,n} d\bar{z}_{1,n} e^{-2A_n z_{1,n}
\bar{z}_{1,n} + \gamma_n(\bar{z}_{1,n}\alpha_n + \bar{z}_{2,n}z_{1,n})}
\nonumber\\
&& \mbox{(which on integrating over the $z_{1,n},\bar{z}_{1,n}$ variables 
becomes )}\nonumber \\ 
&=&\prod_n |{\cal N}'_n|^{2N} (\frac{\pi}{2A_n})
e^{-A_n (\alpha_n \alpha_n^\dagger+\alpha'_n \alpha'^\dagger_n)}
\int dz_{N-1,n}d\bar{z}_{N-1,n} e^{-2A_n z_{N-1,n}
\bar{z}_{N-1,n} + \gamma_n(\alpha'^\dagger_n z_{N-1,n}+ \bar{z}_{N-1,n}
z_{N-2,n})}\times \cdots \nonumber \\
&& \cdots \times \int dz_{2,n} d\bar{z}_{2,n} e^{-2A_n z_{2,n}
\bar{z}_{2,n} + \frac{\gamma_n^2}{2A_n}\bar{z}_{2,n}\alpha_n + 
\gamma_n\bar{z}_{3,n}z_{2,n}} \;\; \nonumber \\
&=& \cdots \nonumber \\
&=&\prod_n |{\cal N}'_n|^{2N}(\frac{\pi}{2A_n})^{N-1}
 e^{-A_n (\alpha_n \alpha_n^\dagger+\alpha'_n
\alpha'^\dagger_n)+ \gamma_n (\frac{\gamma_n}{2A_n})^{N-1}\alpha'^\dagger_n 
\alpha_n},
\label{a.13}
\eea
where  we have repeated used the identity
\be
\int dz d\bar{z} e^{-Az\bar{z}+ bz + c\bar{z}} = \frac{\pi}{A}e^{\frac{bc}{A}}.
\label{a.12}
\ee
The bars here denote complex conjugation ( just as the daggers).

Hence, we can see
\bea
\tr(\bbox{\rho}_{red}^N ) &=& \prod_n \int d\alpha_n d\alpha_n^\dagger 
\bbox{\rho}_{red}^{(n)}(\alpha_n;\alpha_n) \nonumber \\
&=&\prod_n |{\cal N}'_n|^{2N}(\frac{\pi}{2A_n})^{N-1}\frac{\pi}{2A_n 
-\gamma_n (\frac{\gamma_n}{2A_n})^{N-1}} \nonumber \\
&=& \prod_{n}\frac{( 2 A_n - \gamma_n )^N}{
( 2 A_n)^N - (\gamma_n)^N} =
 \prod_{n} \frac{(1 - \frac{|C_n|^2}{4 A_nB_n} )^N}{
1 - (\frac{|C_n|^2}{4 A_nB_n})^N},
\label{a.14}
\eea
where we have used (\ref{a.10}).
This expression is invariant under the simultaneous rescaling
$A_n,B_n,C_n \ra \lambda A_n ,\lambda B_n, \lambda C_n$ of the constants.
Hence so will be the entropy.
Using it in (\ref{a.11}), it readily follows that
\be
S= \sum_n \left[ \frac{1}{(2A_n - \gamma_n)} \left( 2A_n \ln 2A_n -
\gamma_n \ln  \gamma_n \right) - \ln ( 2A_n-\gamma_n ) \right].
\label{a.15}
\ee
The constants $C_n$ are determined by the coupling between the variables
$\{\alpha_n\}$ and $\{a_n\}$. When the couplings vanish,  $C_n$ are zero and
so are $\gamma_n$ and hence, 
as one can see  from (\ref{a.15}), the entanglement entropy vanishes as well.

In the weak coupling limit ($q \ra 0$), the following simplifications occur. 
One can write using equations (\ref{energy},\ref{a.4})
\bea 
C_n &=&-\sqrt{\frac{2\pi \sigma_H}{n t}} \frac{2 i n
\omega_n N_{nm} J'_n(\chi_{nm})}{(\omega_n+\frac{2\sigma_H}{t}+ \frac{2n}{R})} \;,
\nonumber \\
&\ra&-\sqrt{\frac{2\pi k q^2}{n t}} \frac{2i n
\omega_n N_{nm} J'_n(\chi_{nm})}{(\omega_n+ \frac{2n}{R})},
\label{a.16}
\eea
where we have dropped the term proportional to $q^2$ in the denominator as
it would be small compared to the other terms in the weak coupling limit.

Therefore, one finds
\be
\gamma_n =\frac{|C_n|^2}{2B_n}= \frac{4\pi k q^2 R}{ t} \frac{ n
 N^2_m J'^2_n(\chi_{nm}) \chi_n}{(\chi_{n}+ 2n)^2 }, 
\label{a.17}
\ee
where we have used the relation $\chi_{n}= \omega_n R$. Note that
$\gamma_n$ is nothing but ${\bf p}_n$ for a single bulk mode.
 Due to the relation 
$A_n=\frac{1}{2}$, the entropy expression now reduces to 
\be
S= \sum_n \left[ \frac{1}{(1 - \gamma_n)} \left( -
\gamma_n \ln  \gamma_n \right) - \ln (1-\gamma_n ) \right].
\label{a.18}
\ee
 For small $\gamma_n$ this expression reduces to that appearing in equation 
(\ref{6.1}) and hence shows exactly the same scaling behavior. 
\end{document}